\documentclass[11pt, a4paper, oneside]{article}
\usepackage[left=25mm,right=20mm]{geometry}
\usepackage{amssymb}
\usepackage{amsmath}
\usepackage{amsthm}

\usepackage[english]{babel}
\usepackage[utf8]{inputenc}
\usepackage{xcolor}
\usepackage[ruled,vlined]{algorithm2e}
\usepackage{float}
\usepackage{multicol}
\usepackage{booktabs}
\usepackage{graphicx}
\usepackage{colortbl}
\usepackage[hidelinks]{hyperref}
\usepackage{adjustbox}
\usepackage{natbib}
\usepackage{cleveref}
\usepackage{comment}
\usepackage{tikz}
\usetikzlibrary{arrows.meta}
\usetikzlibrary{shapes.geometric}
\usepackage{tikzscale}
\usepackage{arydshln}
\usepackage{tabularx}
\usepackage{mathtools}
\usepackage{subcaption}
\usepackage[normalem]{ulem}
\usepackage{mathrsfs}
\usepackage{orcidlink}
\usepackage{authblk}
\usepackage{longtable}

\usepackage[autostyle, english = american]{csquotes}
\MakeOuterQuote{"}

\newcommand{\EV}[1]{\mathbb{E}_{#1}}
\newcommand{\e}[1]{e_{#1}^{[n]}}
\newcommand{\K}[1]{K_{#1}^{[n]}}
\newcommand{\BarK}[1]{\Bar{K}_{#1}^{[n]}}
\newcommand{\trueP}{\mathbb{P}}
\newcommand{\PH}[1]{\mathcal{P}_0(#1)}

\DeclareMathOperator*{\argmax}{argmax}
\theoremstyle{plain}

\chardef\bslash=`\\ 

\newtheorem{prop}{Proposition}
\newtheorem{lemma}{Lemma}
\newtheorem{defi}{Definition}

\title{Anytime-valid simultaneous lower confidence bounds for the true discovery proportion}
\author{Friederike Preusse}
\date{}
\affil{Institute for Statistics, University of Bremen, Germany\\
	preusse@uni-bremen.de}

\begin{document}
	
	\maketitle

\begin{abstract}
We propose a method that combines the closed testing framework with the concept of safe anytime-valid inference (SAVI) to compute lower confidence bounds for the true discovery proportion in a multiple testing setting. 
The proposed procedure provides confidence bounds that are valid at every observation time point and that are simultaneous for all possible subsets of hypotheses. While the hypotheses are assumed to be fixed over time, the subsets of interest may vary.
Anytime-valid simultaneous confidence bounds allow us to sequentially update the bounds over time and allow for optional stopping. 
This is a desirable property in practical applications such as neuroscience, where data acquisition is very time and cost consuming. 
We also present a computational shortcut which makes the application of the proposed procedure feasible when the number of hypotheses under consideration is large. We illustrate the performance of the proposed method in a simulation study and give some practical guidelines on the implementation of the proposed procedure. Our method is applied to real data from a functional magnetic resonance imaging experiment.
\end{abstract}

\section{Introduction}
When testing multiple hypotheses, researchers might be interested in grouping hypotheses and making inference on these groups of hypotheses. To this end, confidence bounds for the number of false null hypotheses (called true discoveries) in a subset of considered null hypotheses can be utilized.
If the subset of null hypotheses of interest is specified a-priori, procedures to compute lower confidence bounds for the true discovery proportion (TDP) are given in \citet{MeinshausenRice2006, GeLi2012, Patra2016, Preusse2025}, among others.
Simultaneous lower confidence bounds for the TDP make inference on the true discoveries simultaneously for all possible subsets of hypotheses. This allows the practitioner to specify sets of hypotheses of interest based on the corresponding confidence bounds for the TDP. 
Methods to compute simultaneous lower confidence bounds for the TDP have been proposed by \citet{GenoveseWasserman2006,Goeman2011, Hemerik2019,Blanchard2020,Katsevich2020,Vesely2023}, among others, and have been applied to genomics \citep{Ebrahimpoor2020} or functional magnetic resonance imaging (fMRI) \citep{Rosenblatt, Blain2022, Andreella2023}. %
The aforementioned procedures are based on closed testing \citep{Marcus1976}, which has been shown to be the only admissible framework for confidence bounds for the TDP \citep{Goeman2021}. 

These established methods require a fixed sample size; early stopping or increasing the sample size after the data analysis is not allowed. Time and/ or monetary constraints might require stopping the data acquisition and analyzing the available data earlier than planned.
These constraints might be lifted at a later time point and being able to resume the study seems advantageous. Even if there are no constraints, stopping the data gathering process as soon as the results are deemed to be interesting can save both time and money compared to sampling until a pre-specified sample size is achieved. Deciding based on the current results or external rules whether data acquisition should be continued is called optional stopping \citep[c.f.][]{SafeTesting}. Even if the data acquisition is stopped and the data are analyzed, optional stopping allows for the continuation of the data acquisition at a later time point and combining the new data with the previous results. Procedures that are valid under optional stopping are often called anytime-valid procedures.

Sequentially testing a single hypothesis under optional stopping can be done using an e-process, which is an extension of the recently proposed e-value \citep[see e.g.][]{VovkWang2021,SafeTesting, Koolen2022, Ruf2023}.
An e-value can be interpreted as measuring the evidence against a null hypothesis \citep{VovkWang2021}. An e-process is an anytime-valid e-values, that is, an e-value that is valid at any time point of the process. The construction of e-processes is an active topic of research \citep[among others]{Koolen2022, Ramdas2023,Perez2024, terSchure2024,Turner2024,VovkWang2024a, Wang2025}.
Simultaneous lower confidence bounds for the TDP based on e-values have been proposed by \citet{VovkWang2023, VovkWang2024}. Their approach does not allow for updating the confidence bounds over time. \citet{Fischer2024} update the lower confidence bounds for the TDP using e-values. In contrast to our work, they focus on the online multiple testing setting, that is, the hypotheses of interest are updated over time. We study the offline setting, where the set of hypotheses of interest is considered to be fixed.

In this work, we propose a procedure to compute anytime-valid simultaneous lower confidence bounds for the TDP in the offline setting. To this end, we combine the closed testing framework with anytime-valid testing procedures, focusing primarily on e-processes.  Furthermore, a computational shortcut is derived so that computation of the bounds is feasible even if the number of hypotheses of interest is large. 
In contrast to established procedures for simultaneous lower TDP confidence bounds, the proposed simultaneous confidence bounds are anytime-valid and hence the proposed procedure allows for optional stopping. Compared to existing work regarding e-processes, we do not focus on constructing e-processes but integrate e-processes in the closed testing framework to obtain anytime-valid simultaneous TDP bounds in an offline setting.

The rest of the work is structured as follows. In the next section, we specify our framework and review related literature. We propose the procedure used to compute anytime-valid simultaneous confidence bounds for the TDP in Section \ref{sec:theory_ssucb}. In Section \ref{sec:practical_considerations}, we discuss some practical aspects. 
We investigate the behavior of the proposed method in a simulation study in Section \ref{sec:simulations}. The proposed procedure is applied to real data of a fMRI experiment in Section \ref{sec:case_study}. We end with a short discussion in Section \ref{sec:discussion}. Proofs are deferred to the appendix.

\section{Notation and Related Literature}
\label{sec:preliminaries_and_related_literature}
First, notation is introduced and then a brief review of simultaneous confidence bounds for the TDP based on closed testing is given. This section concludes with a short overview of e-values and e-processes.

In the following, we use bold letters to refer to vectors or, if explicitly stated, matrices. Furthermore, when we write hypothesis we are referring to a null hypothesis, unless explicitly stated otherwise.
We consider the probability space $(\Omega, \mathcal{F}, \mathcal{P})$, 
where $\mathcal{P}$ is the set of all probability measures on $(\Omega,\mathcal{F})$.
A probability distribution is denoted by $\mathbb{P}$ with corresponding expected value $\mathbb{E}_\mathbb{P}$. 
We aim to make inference on the set of hypotheses $\mathcal{H}\coloneq\{H_1,\ldots, H_m\}$ and refer to $H_i$, $i\in\{1,\ldots,m\}$, as elementary hypothesis. The considered hypotheses are composite hypotheses, that is, each elementary hypothesis $H_i$ corresponds to a set of probability measures $\mathcal{P}_0(i)\subset\mathcal{P}$. 
The hypothesis $H_i$ is true under an arbitrary but fixed $\trueP\in\mathcal{P}$ if and only if $\mathbb{P}\in\PH{i}$.

Let $\mathcal{C}=2^{\{1,\ldots,m\}}\setminus \emptyset$ denote the collection of all nonempty subsets of $\{1,\ldots,m\}$.
Intersection hypotheses are denoted by $H_I\coloneq\bigcap_{i\in I}H_i$, where $ I\in\mathcal{C}$ is the set of indices corresponding to the elementary hypotheses intersected by $H_I$. The size of the set $I$ is denoted by $|I|$, with $|I|\geq 1$. Therefore, $H_I$ might refer to an elementary or intersection hypothesis, while $H_i$ always denotes an elementary hypothesis. We denote by $\PH{I}\coloneq \bigcap_{i\in I}\PH{i}$ the set of all probability measures on $(\Omega, \mathcal{F})$ corresponding to $H_I$.
The set $T(\mathbb{P})\equiv T\coloneq\{i\in\{1,\ldots,m\}|\trueP\in \PH{i}\}$ refers to the index set of all true elementary hypotheses under some $\trueP\in\mathcal{P}$. 
Since $H_T$ intersects all elementary hypotheses that are true under $\trueP$, $\trueP\notin\PH{i}$ for every $i\notin T$. 
Thus, an intersection hypothesis $H_I$ is true under $\trueP\in\mathcal{P}$ if and only if $I\subseteq T$. 

The set of indices of the rejected elementary hypotheses is denoted by $R$, with $R\in\mathcal{C}$. We do not specify any rejection rule to determine $R$. Indeed, the rejected hypotheses correspond to (a subset of) the elementary hypotheses that are of interest to the practitioner. Every rejected elementary hypothesis is considered to be a discovery and a true, but falsely rejected elementary hypothesis corresponds to a false discovery. Because $R$ specifies the set of discoveries, we call it a discovery set. 
Denote by $T_R(\trueP)\equiv T_R\coloneq T\cap R$ the set of indices corresponding to discoveries which are false under some $\trueP\in\mathcal{P}$. 
The number of false discoveries is denoted by $|T_R|=\tau(R)$. The false discovery proportion (FDP) is given by $\pi_0(R)=\tau(R)/|R|$ and the true discovery proportion (TDP) is given by $\pi_1(R)=1-\pi_0(R)$. 

Lastly, let $\phi_\alpha$ denote a family of local level-$\alpha$ test with binary outcome which rejects hypothesis $H_I$ iff $\phi_\alpha^I=1$ and which controls the local type I error rate at level $\alpha$, i.e., $\mathbb{P}(\phi_\alpha^I=1)\leq\alpha$ for all $\mathbb{P}\in\PH{I}$ and any $\alpha\in(0,1)$. 
\subsection{Simultaneous lower confidence bounds for the TDP}
In this section, we provide more details about the construction of simultaneous lower confidence bounds for the TDP.
Since $|R|$ is observable, $\tau(R)$ is the only unknown quantity in $\pi_0(R)$ and $\pi_1(R)$. Thus, to compute lower confidence bounds for $\pi_1(R)$, it is sufficient to compute upper confidence bounds for $\tau(R)$, which we will focus on in the following. 
\citet{GenoveseWasserman2006} and \citet{Goeman2011} propose simultaneous upper confidence bounds for $\tau(R)$ based on closed testing. \citet{Goeman2021} show that methods to compute confidence bounds for $\tau(R)$ are either based on closed testing or uniformly improvable by closed testing.
Closed testing procedures are a family of multiple testing methods proposed by \citet{Marcus1976}. These procedures can be used to control the family wise error rate, i.e., controlling the probability of making at least one false rejection. Denote by $\mathcal{X}_\alpha$ the index set of the hypotheses rejected by closed testing. Let $\mathcal{U}_\alpha$ be the index set of the hypotheses that are rejected by a given family of local level-$\alpha$ test $\phi_\alpha$. 
The intersection hypothesis $H_I$ is rejected by closed testing based on $\phi_\alpha$, i.e., $I\in\mathcal{X}_\alpha$, if and only if for every $J\supseteq I$, $J\in\mathcal{U}_\alpha$. In other words, closed testing rejects the hypothesis $H_I$ if and only if every hypothesis intersecting $H_I$ is rejected by a given local level-$\alpha$ test. Both sets $\mathcal{U}_\alpha$ and $\mathcal{X}_\alpha$ depend on the choice of the local level-$\alpha$ tests. In the following, we consider $\phi_\alpha$ to be fixed and thus omit this dependency in the notation to improve readability. 

\citet{Goeman2011} define \begin{equation}
	\label{eq:simple_bounds}
	c_\alpha(R)\coloneq\max\{|I|:I\subseteq R, I\neq\emptyset, I\notin\mathcal{X}_\alpha\} 
\end{equation} if the maximum exists and $c_\alpha(R)\coloneq0$ otherwise. That is, $c_\alpha(R)$ corresponds to the size of the largest hypothesis which only intersects discoveries and is rejected by closed testing. They show that $c_\alpha(R)$ is a simultaneous $1-\alpha$ upper confidence bound for $\tau(R)$, i.e., \begin{equation*}
	\label{eq:prob_simple_bounds}
	\trueP\left(\tau(R)\leq c_\alpha(R) \quad \forall R\in\mathcal{C}\right)\geq 1-\alpha, \quad \text{for all } \trueP\in\mathcal{P}\text{ and }\alpha\in(0,1).
\end{equation*}

The proof is based on the observation that $\tau(R)> c_\alpha(R)$ if and only if the largest true hypothesis in the discovery set, $H_{T_R}$, is rejected by closed testing. 
Under any $\trueP\in\mathcal{P}$, $H_{T_R}$ is true by definition and the
probability that $H_{T_R}$ is rejected by closed testing is less than or equal to $\alpha$. 
This is because the event $T_R\in\mathcal{X}_\alpha$ requires that $H_T$ is rejected by $\phi_\alpha$, which has a probability of less than or equal to $\alpha$ by definition of $H_T$ as the largest true hypothesis under $\trueP\in\mathcal{P}$. 
Since the event $T\in\mathcal{U}_\alpha$, i.e., $H_T$ being rejected by $\phi_\alpha$, does not depend on the discovery set $R$, the bounds are simultaneously valid for all $R\in\mathcal{C}$.

This approach requires testing up to $2^m-1$ hypotheses. When the utilized local tests are based on p-values, \citet{Goeman2019} propose a computational shortcut which reduces the number of hypotheses that need to be tested explicitly. 

This procedure to compute simultaneous upper confidence bounds for $\tau(R)$ has been extended in recent years, using for example sum tests \citep{Vesely2023} or permutation tests \citep{Hemerik2019} as local level-$\alpha$ tests. The approach has also been adapted to different settings, such as knock-offs \citep{Li2024}. 
\citet{VovkWang2023} extend the approach by \cite{Goeman2011} to e-values, which are discussed in the following.

\subsection{E-variables and e-processes}
\label{subsec:e-values}
First, we give a formal definition of an e-variable. 
\begin{defi}
	An e-variable $E_I$ corresponding to hypothesis $H_I$ is a nonnegative random variable with $\EV{\mathbb{P}}[E_I]\leq 1$ for all $\mathbb{P}\in\PH{I}$.
\end{defi}
The realization of $E_I$ is called an e-value. 
Because of Markov's inequality, $\mathbb{P}(E_I\geq 1/\alpha)\leq \alpha$ for all $\mathbb{P}\in \PH{I}$ and $\alpha\in(0,1)$. A local level-$\alpha$ test is thus given by $\phi_\alpha^{I}(E)\coloneq\mathbb{I}\{E_I\geq 1/\alpha\}$, where $\mathbb{I}(\cdot)$ denotes the indicator function \citep{SafeTesting, VovkWang2021}. 
Simultaneous upper confidence bounds for $\tau(R)$ based on closed testing utilizing the family of local level-$\alpha$ test $\phi_\alpha(E)$ are discussed in \citet{VovkWang2023} and \citet{VovkWang2024}. They determining $c_\alpha(R)$ so that there is strong evidence that $\tau(R)\leq c_\alpha(R)$. \citet{VovkWang2023, VovkWang2024} evaluate the strength of the evidence against the null hypothesis based on the corresponding e-value. Their approach allows for more lenient critical values of the applied local tests than $\phi_\alpha(E)$, but it does not offer the same coverage guarantees as as the approach by \citet{Goeman2011}.

There are many different families of e-variables. These families are often based on a likelihood ratio or Bayes factor \citep[c.f.][]{SafeTesting, Wasserman2020}. 
There are some guidelines on how to choose an e-variable in a specific setting. Mainly, the e-variable should grow as fast as possible if the evidence against the corresponding null hypothesis increases. \citet{SafeTesting} introduce the growth-rate optimal in the worst-case (GROW) criterion. This criterion measures how fast an e-variable can accumulate evidence against the null hypothesis under the worst-case alternative. It can be used to compare families of e-variables or specify e-variables. Furthermore, it is possible to calibrate e-variables based on p-variables and vice-versa \citep{VovkWang2021}. Remember, that a valid p-variable $P_I\in[0,1]$ corresponding to hypothesis $H_I$ is a random variable with $\mathbb{P}(P_I\leq\alpha)\leq\alpha$ for all $\mathbb{P}\in\PH{I}$. Most notable, $P_I^{(E)}\coloneq\min(1/E_I,1)$ is a valid, albeit conservative p-variable \citep{VovkWang2021,SafeTesting}. However, the inverse of a p-variable does not need to be a valid e-variable. 

In contrast to p-variables, merging e-variables is rather straightforward.
Functions to merge e-variables, so called e-merging functions, have been studied by \citet{VovkWang2021}. 
The e-variable corresponding to the intersection hypothesis $H_I$ can be defined using the arithmetic mean as e-merging function, such that $E_I=\sum_{i\in I}E_i /|I|$. Indeed, under arbitrary dependence between the e-variables, utilizing a (weighted) arithmetic mean is the only admissible way of merging e-variables \citep{Wang2025a}. That is, there exists no other e-merging function that returns e-variables which are uniformly larger than the e-variables returned by (weighted) arithmetic averages. 

In our approach, we are interested in making inference on a set of hypotheses which is considered to be fixed at multiple observation time points, while allowing for optional stopping.  
This can be done using e-processes. Let $(\mathcal{F}_n)_{n\geq 0}$ be a filtration, i.e., a sequence of sigma-algebras $\mathcal{F}_0\subseteq\mathcal{F}_1\subseteq\ldots\subseteq\mathcal{F}$. 

\begin{defi}
	An e-process corresponding to $H_I$ is a nonnegative process $(E_I^{[n]})_{n\geq 0}$ adapted to some filtration $(\mathcal{F}_n)_{n\geq 0}$ with $\EV{\mathbb{P}}[E^{[\nu]}_I]\leq 1$ for any $\mathcal{F}$-stopping time $\nu$ and for all $\mathbb{P}\in \PH{I}$. \citep{Ruf2023} 
\end{defi}
At any $\mathcal{F}$-stopping time $\nu$, $E_I^{[\nu]}$ is a valid e-variable for $H_I$. An e-process variable is also called safe anytime-valid e-variable \citep[c.f.][]{Koolen2022}. We denote the set of all $\mathcal{F}$-stopping times by $\mathfrak{N}$. 
If $H_I$ is true, the probability that a corresponding e-process
ever exceeds a given threshold is bounded from above.
Specifically, \citet{Ruf2023} show that \begin{equation}
	\label{eq:Villes_inequality}
	\mathbb{P}(\exists \nu\in\mathfrak{N}: E_I^{[\nu]}\geq \frac{1}{\alpha})\leq \alpha \quad \text{for every } \alpha\in[0,1]\text{ and every }\mathbb{P}\in\PH{I}.
\end{equation}

This can be used to define anytime-valid local level-$\alpha$ tests. We say that a family of local level-$\alpha$ tests $\phi_\alpha^{[\nu]}$ is anytime-valid if $\mathbb{P}(\exists \nu\in\mathfrak{N}:\phi_\alpha^{[I,\nu]}=1)\leq \alpha$ for all $\mathbb{P}\in\PH{I}$. That is, the probability that a true hypothesis is ever rejected by an anytime-valid local level-$\alpha$ test is at most $\alpha$. 
An anytime-valid local level-$\alpha$ test based on an e-process is given by  $\phi_\alpha^{[I,\nu]}(E)\coloneq\mathbb{I}\{E_I^{[\nu]}\geq 1/\alpha\}$, where $E_I^{[\nu]}$ denotes the value of the e-process at time $\nu\in\mathfrak{N}$. Generally, there are no guarantees that an e-process is non-decreasing if the corresponding hypothesis is false. Therefore, $\phi_\alpha^{[\nu]}(E)$ might not reject an hypothesis at a given time point $\nu+k$, $\nu+k\in\mathfrak{N}$, even if it rejected the hypothesis at time $\nu$. The running maximum of any anytime-valid local level-$\alpha$ test, $\tilde{\phi}_\alpha^{[I,\nu]}\coloneq \max_{\ell\leq \nu}\phi_\alpha^{[I,\nu]}$, is an anytime-valid local level-$\alpha$ test as well \citep{Ramdas2022}. An hypothesis $H_I$ rejected by $\tilde{\phi}_\alpha^{[\nu]}$ at time $\nu$ is rejected by this test at every following time point as well. Utilizing $\tilde{\phi}_\alpha^{[n]}$ ensures carefree testing in the sense of \citet{Tavyrikov2025}, that is, no disadvantage arises for the practitioner from gathering more data.
 
As with e-variables, there are several families of e-processes and many of them are based on likelihood ratios \citep[c.f.][]{ Koolen2022,Perez2024,terSchure2024, Wang2025}. Alternatively, e-processes can be constructed by merging e-values \citep{VovkWang2024a}. An overview of different families of e-processes is given in \citet{Ramdas2023}. \citet{Choe2025} show that any e-process defined with respect to the filtration $(\mathcal{F}_n)_{n\geq0}$ can be adapted to be a valid e-process with respect to another filtration $(\mathcal{G}_n)_{n\geq0}$.

In the next section, we combine simultaneous confidence bounds based on closed testing with anytime-valid local level-$\alpha$ tests and propose a procedure to compute anytime-valid upper confidence bounds for $\tau(R)$.

		\section{Anytime-valid simultaneous confidence bounds} 
\label{sec:theory_ssucb}
We consider the filtered probability space $(\Omega, \mathcal{F},(\mathcal{F}_n)_{n\in\mathbb{N}_0}, \mathcal{P})$, with $\Omega$, $\mathcal{F}$ and $\mathcal{P}$ as in Section \ref{sec:preliminaries_and_related_literature}. Without loss of generality, let $(\mathcal{F}_n)_{n\in\mathbb{N}_0}$ be the natural filtration corresponding to the observational units. Note that we consider a global natural filtration, that is, the natural filtration for each data stream per hypothesis is considered to be the same.
 We make no assumptions regarding the dependency structure of the data.
At each observation time point $n\in\mathbb{N}_0$, we consider the set of hypotheses $\mathcal{H}$. 

Denote by $\mathcal{U}_\alpha^{[n]}$ the index set of the hypotheses rejected by a given family of anytime-valid local level-$\alpha$ tests at time $n$. 
Furthermore, let $\mathcal{X}_\alpha^{[n]}$ be the index set corresponding to the hypotheses rejected by closed testing based on $\mathcal{U}_\alpha^{[n]}$ at time $n$. Once again, we notationally omit the dependency of the sets $\mathcal{U}_\alpha^{[n]}$ and $\mathcal{X}_\alpha^{[n]}$ on the family of anytime-valid local level-$\alpha$ tests since this family is considered to be fixed.

Our main result is stated in Proposition \ref{prop_1}, which defines anytime-valid simultaneous upper confidence bounds for the number of false discoveries. These bounds are an extension of the bounds given in Eq.~\eqref{eq:simple_bounds}. At each time point $n$, the size of the largest hypothesis that only intersects discoveries and that is rejected by closed testing at that time $n$ is considered. The proposed anytime-valid simultaneous upper confidence bounds at time $n$ are the minimum of these values observed up to time $n$.

\begin{prop}
	\label{prop_1}
	Define $c_{\alpha}^{[0]}(R)=|R|$ and for all $n\in\mathbb{N}$ \begin{equation*}
		c_{\alpha}^{[n]}(R)\coloneq\max\{|I|:I \subseteq R, I\neq\emptyset, I\notin \mathcal{X}_\alpha^{[n]}\},
	\end{equation*} if this maximum exists and $c_{\alpha}^{[n]}(R)\coloneq 0$ otherwise. Furthermore, define for all $n\in\mathbb{N}$\begin{equation}
		\label{eq:savi_bound}
	\tilde{c}_\alpha^{[n]}(R)\coloneq\min_{0\leq\ell\leq n}\{c_\alpha^{[\ell]}\}.
		\end{equation}
	Then,\begin{equation*}
		\label{eq:savi_bound_prob}
		\trueP\left(\tau(R)\leq \tilde{c}_\alpha^{[n]}(R) \quad \forall R\in\mathcal{C}, \forall n\in\mathbb{N}_0\right)\geq 1-\alpha,
	\end{equation*} for all $\trueP\in\mathcal{P}$ and $\alpha\in(0,1)$. \\
\end{prop}
The proof is provided in Appendix A.1. The upper bound $\tilde{c}_\alpha^{[n]}$ is carefree in the sense that it is non-increasing over time, regardless of the utilized anytime-local level-$\alpha$ tests determining $\mathcal{U}_\alpha^{[n]}$.
Based on Proposition \ref{prop_1}, an anytime-valid simultaneous lower confidence bound for the TDP in the discovery set $R$ is computed as $\tilde{d}_\alpha^{[n]}(R)=1-\tilde{c}_{\alpha}^{[n]}(R)/|R|$.

Proposition \ref{prop_1} is valid for any family of anytime-valid local level-$\alpha$ tests determining $\mathcal{U}_\alpha^{[n]}$. E-processes play an essential role for anytime-valid inference \citep[see e.g.][]{Ramdas2022}. Hence, we restrict the following investigations to the family of anytime-valid local level-$\alpha$ test $\phi_\alpha^{[n]}(E)$, as defined at the end of Section \ref{subsec:e-values}. 
If $\phi_\alpha^{[n]}(E)$ is utilized as family of anytime-valid local level-$\alpha$ tests, valid e-processes for every elementary hypothesis $H_i$, $i\in\{1,\ldots,m\}$ are required. In this work, we only consider e-processes with respect to the global natural filtration, hence all e-processes are defined on the same filtration. 
This allows us to compute confidence bounds after every observation. 
Based on the e-processes corresponding to the elementary hypotheses, the e-processes corresponding to intersection hypotheses can be computed. It has been pointed out by \citet{Ramdas2023} that e-processes of elementary hypotheses can be merged, similarly to e-values. Since we allow for arbitrary dependence between the e-processes, we consider only the arithmetic mean as e-merging function and thus compute the e-process at time $n$ corresponding to the intersection hypothesis $H_I$ as $E_I^{[n]}=\sum_{i\in I}E_{i}^{[n]}/|I|$.

Computing $\tilde{c}_{\alpha}^{[n]}(R)$ at time point $n$ requires testing up to $2^m-1$ hypotheses. Computational shortcuts can reduce the number of explicitly tested hypotheses. We propose such a shortcut when utilizing $\phi_\alpha^{[n]}(E)$ and the arithmetic mean as e-merging function. We assume that the discovery set $R$ is fixed over time. Lemma \ref{lemma_shortcut} introduces an alternative formulation of the carefree anytime-valid simultaneous lower confidence bounds for the number of false discoveries compared to that in Proposition \ref{prop_1} when $\phi_\alpha^{[n]}(E)$ and the arithmetic mean e-merging function is used.

\begin{lemma}
	\label{lemma_shortcut}
	Assume that the arithmetic mean is utilized as e-merging function and that $\phi_\alpha^{[n]}(E)$ is chosen as family of anytime-valid local level-$\alpha$ tests.\\
	Denote by $\K{h}\subseteq R$, the set of indices of the elementary hypotheses in a discovery set $R$ with the $h$ smallest corresponding e-process values at time point $n$, $n\in\mathbb{N}_0$. Denote by $\BarK{k}\in\mathcal{C}$, $\BarK{k}\cap R=\emptyset$, the set of indices corresponding to the elementary hypotheses with the $k$ smallest e-process values at time point $n$ that are not in the discovery set. \\
	Furthermore, define $k^\ast\coloneq \argmax_k\{k/\alpha-\sum_{j\in\BarK{k}}\e{j}\}$, $0\leq k\leq m-|R|$, with $\sum_{j\in\BarK{0}}\e{j}\coloneq 0$ by convention. \\
	Then $\tilde{c}_\alpha^{[n]}(R)=\tilde{h}(n)$, where $\tilde{c}_\alpha^{[n]}(R)$ is defined in Proposition \ref{prop_1}, Eq.~\eqref{eq:savi_bound},  $\tilde{h}(0)=|R|$ and for all $n\in\mathbb{N}$
	\begin{equation}
		\label{eq:shortcut}
	\tilde{h}(n)\coloneq\max_{1\leq h\leq \tilde{h}(n-1)}\left\{h:\sum_{i\in \K{h}}\e{i}-\frac{h}{\alpha} < \frac{k^\ast}{\alpha}-\sum_{j\in\BarK{k^\ast}}\e{j}\right\}
	\end{equation} if this maximum exists and $\tilde{h}(n)\coloneq0$ otherwise. 
\end{lemma}
The proof is provided in Appendix A.2.
For a given time point $n$, Lemma \ref{lemma_shortcut} considers for each possible $h\in\{1,\ldots,\tilde{h}(n-1)\}$ the hypothesis with index set $\K{h}$ of size $h$ which intersects only discoveries. This hypothesis is the least likely to be rejected by closed testing based on $\phi_\alpha^{[n]}(E)$ out of all hypotheses of size $h$ intersecting only discoveries. Lemma \ref{lemma_shortcut} states that the hypothesis with index set $\K{h}$ is not rejected by closed testing if the inequality in Eq.~\eqref{eq:shortcut} holds. In turn, if Eq.~\eqref{eq:shortcut} does not hold, all hypothesis of size $h$ intersecting only discoveries are rejected by closed testing.
To compute $\tilde{c}_\alpha^{[n]}(R)$ as proposed in Lemma \ref{lemma_shortcut}, we iteratively decrease $h\in\{\tilde{h}(n-1),\ldots,1\}$ until the inequality in Eq.~\eqref{eq:shortcut} holds for the first time. Therefore, for a fixed discovery set $R$, at most $|R|$ hypotheses at time point $n$ need to be tested explicitly. Furthermore, notice that the right side of the inequality in Eq.~\eqref{eq:shortcut} does not depend on $h$ and therefore only needs to be computed once for a fixed discovery set at time $n$.
An algorithm to compute $\tilde{h}(n)$ for several discovery sets is provided in Appendix A.3.

	\section{Practical considerations}
\label{sec:practical_considerations}
In this section, we consider the implementation of the proposed anytime-valid simultaneous confidence bounds. 
An important aspect in the practical application of the proposed procedure is the choice of the e-process used to define the anytime-valid local level-$\alpha$ test. 

There are four main aspects which need to be considered when choosing an e-process in practice, namely the validity of the e-process, the growth-rate of the e-process, the choice of the filtration and the computational effort.

First, the proposed procedure based on e-processes is valid as long as the utilized e-process is valid. Thus, the e-process needs to be valid given the dependency structure of the considered data as well as the null distribution. Choosing the e-process also includes considering the available information about the distribution of the data, for example, if the variance-covariance matrix is known or not. This is similar to choosing an appropriate test statistic when working with p-values. 

Second, the faster the utilized e-process can gather evidence against the null hypothesis, the faster the confidence bounds become tight. Therefore, we aim to use an e-process which gathers evidence as fast as possible. The GROW criterion can be applied to e-processes as well. However, while e-processes fulfilling the GROW criterion have the optimal growth-rate in the worst case scenario, they might have a suboptimal growth-rate if the worst-case scenario does not occur. Therefore, in practice, an e-process which has a (comparatively) large growth-rate in multiple scenarios is desirable. 

Third, the set of possible stopping times of the utilized e-process needs to be taken into account. In this work, we only consider e-processes that are defined on the same, global natural filtration. However, for some applications, other filtrations (coarser or finer than the natural filtration) might be more suited. \citet{Choe2025} show how to adjust an e-process to other filtrations than the one the e-process has originally been adapted to. 

Lastly, the computational effort regarding e-processes is of interest in practice. Computing anytime-valid simultaneous upper confidence bounds for the number of false discoveries requires computing e-processes for all of the $m$ elementary hypotheses. The number of considered hypotheses $m$ can be very large. For example, in fMRI more than $100,000$ hypotheses are of interest. Hence, the computational effort, and therefore the time needed to compute a single e-process, should be reasonably small.

An in-depth comparison of e-processes regarding these four aspects is outside the scope of this paper. \citet{Ly2024} compare four e-processes in terms of their growth-rate in different scenarios. These e-processes correspond to the likelihood ratio of a t-test statistic, a test statistic often applied in practice. All four compared e-processes are based on the work by \citet{Gronau2020} and \citet{Perez2024}. The family of e-processes is defined by
\begin{equation}
	\label{eq:Perez2024_e-process}
	E^{[n]}(t)\coloneq\frac{\mathcal{L}(t|\sqrt{n_t}\delta,\lambda)}{\mathcal{L}(t|0,\lambda)},
\end{equation}
where $\mathcal{L}(\cdot|\mu, \lambda)$ is the likelihood of the t-distribution with $\lambda$ degrees of freedom and non-centrality parameter $\mu$. Here, $n_t$ is the effective sample size used to compute the t-test statistic $t$. Furthermore, $\delta$ is a parameter that is chosen such that $E^{[n]}(t)$ is an e-process. 
\citet{Ly2024} recommend using a non-local moment prior on $\delta$, which is a two-bump prior with bumps at $-\delta_{min}$ and $\delta_{min}$, where $\delta_{min}$ is the minimal relevant effect size. They name the resulting e-process the "mom" e-process. While the mom e-process does not necessarily have the largest growth-rate in all scenarios, it's growth-rate appears to be stable across different scenarios. We therefore use the mom e-process in the following simulation study and in the case study.

\section{Simulations}
\label{sec:simulations}
In this section, we investigate the behavior of the proposed procedure in a simulation study. Unlike in the previous sections, we focus on lower confidence bounds for the TDP, $\tilde{d}_\alpha^{[n]}(R)$. These bounds are computed based on the upper confidence bounds for the number of false discoveries, i.e., $\tilde{d}_\alpha^{[n]}(R)=1-\tilde{c}_\alpha^{[n]}(R)/R$.
 In our simulated experiment, $m=1,000$ hypotheses have been tested based on the data of up to $N=100$ subjects. We have simulated sequentially observed data, where data from one subject is observed at each time point. This is in accordance with fMRI studies, where subjects need to be scanned one after the other. 
The data have been independent and identically distributed across the $N$ subjects. The simulated data for one subject $j$ is denoted by $\boldsymbol{y}_j=(y_{j,1},\ldots,y_{j,m})^\top$ and $\boldsymbol{y}_j\sim\mathcal{N}_m(\boldsymbol{\mu},\boldsymbol{V})$ with \begin{equation*}
	\boldsymbol{\mu}_i=\begin{cases}
		0 \qquad \text{if }  i\in\{1,\ldots,500\}\\
		\mu \qquad\text{else},
	\end{cases} \qquad
	\boldsymbol{V}_{a,b}=\begin{cases}
		1 \qquad\text{if }a=b,\\
		\rho\qquad \text{else.}
	\end{cases}
\end{equation*}
We have set $\mu\in\{0.5,1,1.5\}$ and $\rho\in\{0.2,0.6\}$. 
The hypotheses $H_1: \boldsymbol{\mu}_1\leq0,\ldots,H_m:\boldsymbol{\mu}_m\leq0$ have been tested. 
The simulation study has been carried out using the programming language {\tt{R}} \citep{R}. The {\tt{R}} code for the simulation study is available at {\tt{https://github.com/fpreusse/AVARI}}.

For each elementary hypotheses the mom e-process has been computed as implemented in the {\tt{R}} package {\tt{safestats}} \citep{safestats}, with minimal relevant effect size $\delta_{min}=0.5$. The closed testing procedure has been based on the family of anytime-valid local level-$\alpha$ test $\phi_\alpha^{[n]}(E)$. 
The proposed procedure is compared to the established all-resolution inference (ARI) procedure by \citet{Goeman2019} based on p-values, as implemented in the {\tt{R}} package {\tt{hommel}} \citep{hommel}. This procedure returns simultaneous confidence bounds but does not guarantee anytime-validity. Furthermore, this procedure requires that the considered p-values fulfill the positive regression dependency on a subset (PRDS) criterion \citep[Section 3.1]{BenjaminiYekutieli}. Since the simulated data are equi-correlated with positive correlation coefficient, this requirement is fulfilled.

We have considered six different discovery sets with a TDP of $\pi_1\in\{0.1,0.5,0.9\}$ and size $|R|\in\{50, 500\}$. For each discovery set we have computed the $1-\alpha$ anytime-valid simultaneous lower confidence bounds for the TDP, $\tilde{d}_\alpha^{[n]}(R)$, with $\alpha=0.2$. The bounds have been based on a sequentially increasing sample size $11\leq n\leq N$. The following results are based on $1,000$ Monte-Carlo iterations.

The empirical validity of the bounds is examined first. If the empirical non-coverage rate, i.e., $|\pi_1(R)\notin[\tilde{d}_\alpha^{[n]}(R),1]|/1000$, is at most $\alpha$, the bounds are considered valid. At every observation time point the proposed procedure has returned valid bounds, regardless of the level of dependency $\rho$, the size of the discovery set or the effect size $\mu$. This is also true for the bounds based on the ARI procedure, even though these bounds are not guaranteed to be anytime-valid.

\begin{figure}[htbp]
	\centering
	
	\begin{minipage}[b]{0.48\linewidth}
		\centering
		\includegraphics[width=\linewidth]{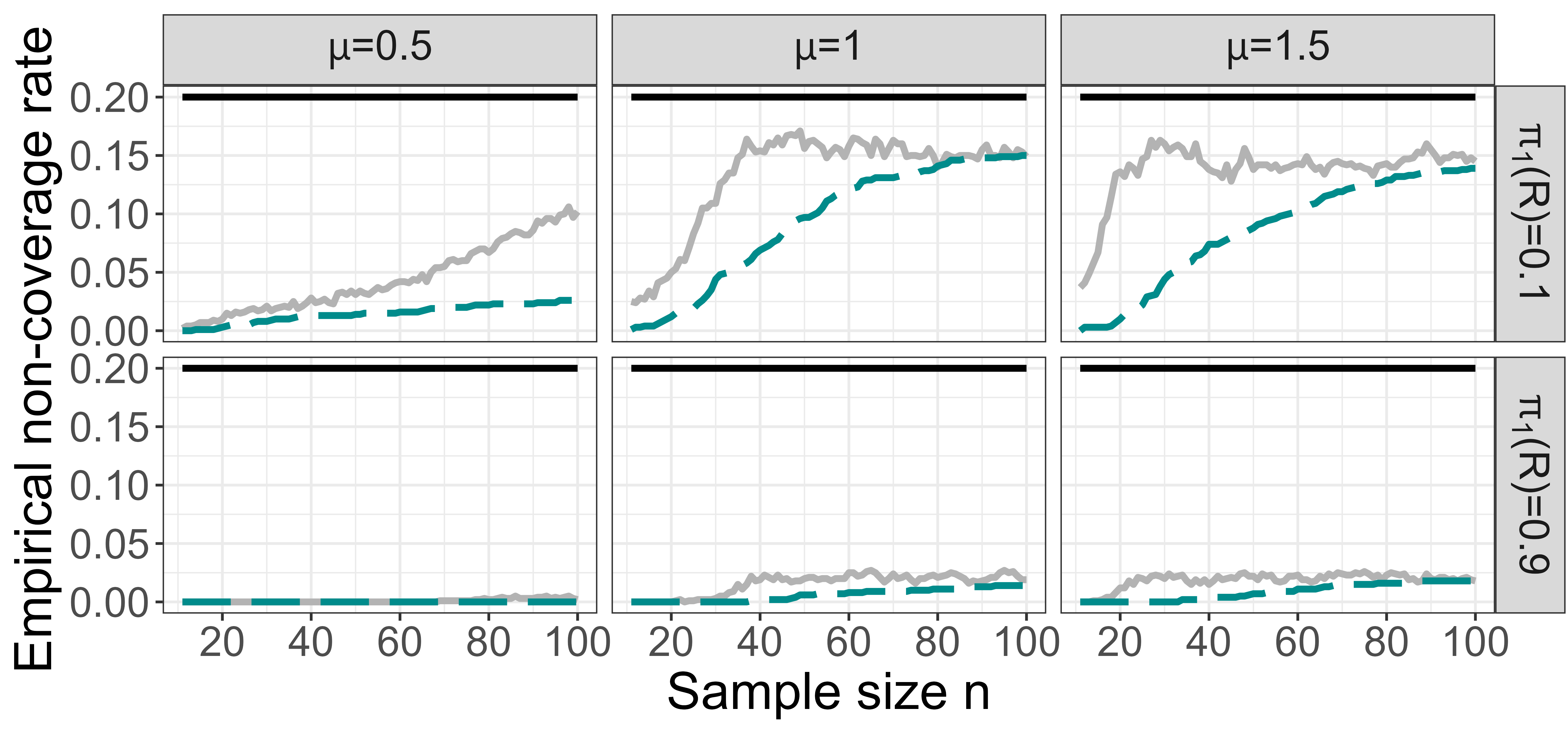}
		\\[3pt]
		\text{$\rho=0.2$, $|R|=500$}
	\end{minipage}%
	\hfill
	\begin{minipage}[b]{0.48\linewidth}
		\centering
		\includegraphics[width=0.94\linewidth]{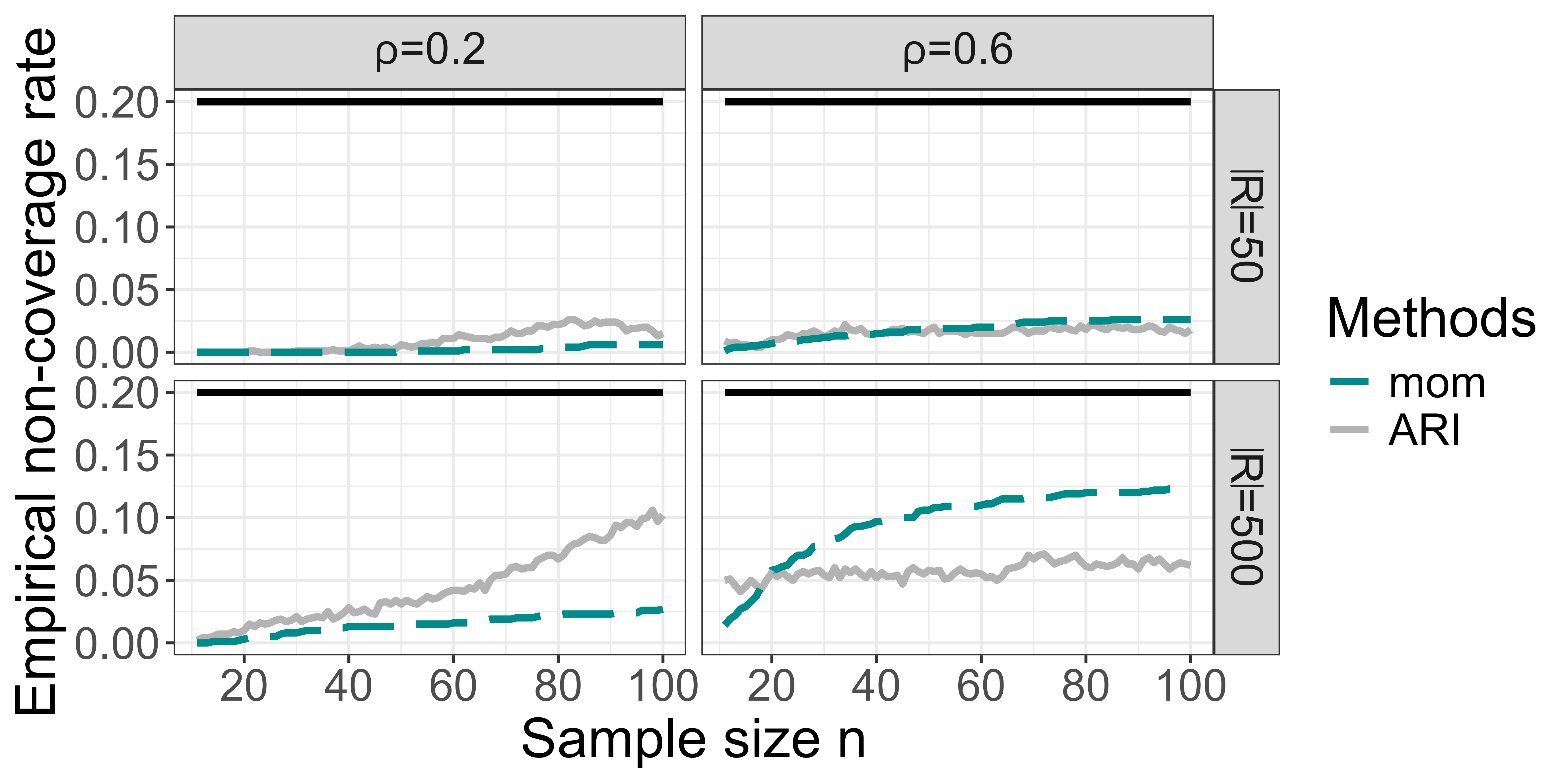}
		\\[3pt]
		\text{$\mu=0.5$, $\pi_1(R)=0.1$}
	\end{minipage}
	
	\caption{Empirical non-coverage rate of the $1-\alpha$ lower confidence bounds for the true discovery proportion for sequentially increasing sample size and different effect sizes $\mu$, true discovery proportion $\pi_1(R)$, level of dependency $\rho$ and size of the discovery set $|R|$. The anytime-valid simultaneous confidence bounds are based on the mom e-process (mom, dark dashed line).  
	The proposed procedure is compared to the all resolution inference procedure (ARI, light gray solid line) by \citet{Goeman2019}.
	The black solid line corresponds to the significance level $\alpha=0.2$. Results are based on $1,000$ iterations.}
	\label{figure:Validity_rho_02}
\end{figure}

The left plot in Figure~\ref{figure:Validity_rho_02} displays the empirical non-coverage rates for sequentially increasing number of observations for $\rho=0.2$, $|R|=500$, $\pi_1(R)\in\{0.1,0.9\}$ and $\mu\in\{0.5,1,1.5\}$. For small $\pi_1(R)$, the empirical non-coverage rate has been larger than for large $\pi_1(R)$. When increasing the effect size from $\mu=0.5$ to $\mu=1$, the empirical non-coverage rate has increased as well. Increasing the effect size from $\mu=1$ to $\mu=1.5$ lead to almost no changes in the empirical non-coverage rate. 
Compared to the proposed procedure, the ARI procedure has returned the larger empirical non-coverage rates, yet the observed bounds have been valid.
On the right side of Figure~\ref{figure:Validity_rho_02} the empirical non-coverage rates are displayed for varying $\rho$ and varying size of the discovery set $|R|$ and fixed effect size $\mu=0.5$ and fixed true discovery rate $\pi_1(R)=0.1$.
When decreasing the size of the discovery set $|R|$, the empirical non-coverage rate has decreased as well. Stronger dependency between the data of the hypotheses ($\rho=0.6$) has lead to larger empirical non-coverage rates for the proposed bounds 
and smaller rates for the ARI procedure compared to lower dependency ($\rho=0.2$).

\begin{figure}[htbp]
	\centering
	
	\begin{minipage}[b]{0.48\linewidth}
		\centering
		\includegraphics[width=\linewidth]{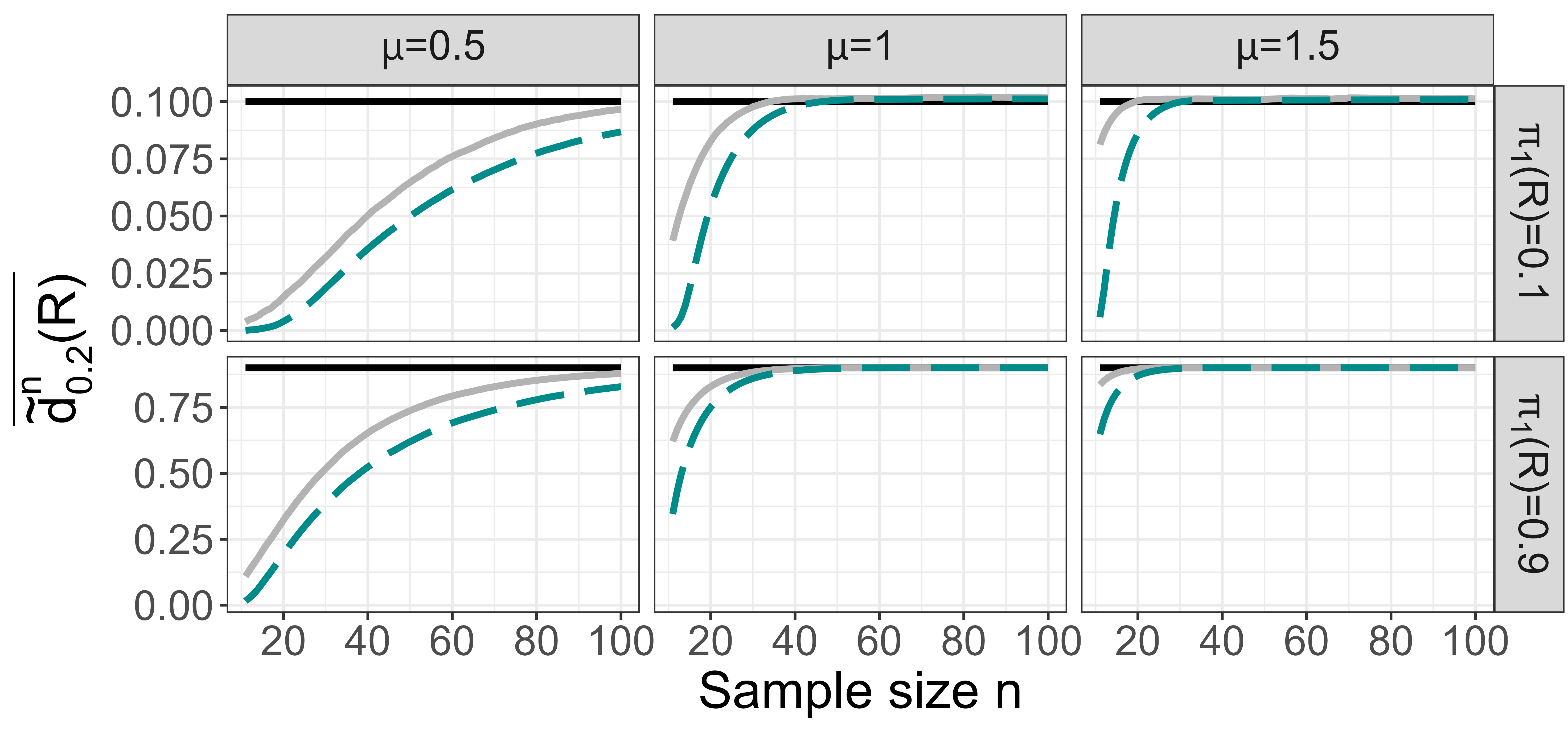}
		\\[3pt]
		\text{$\rho=0.2$, $|R|=500$}
	\end{minipage}%
	\hfill
	\begin{minipage}[b]{0.48\linewidth}
		\centering
		\includegraphics[width=0.935\linewidth]{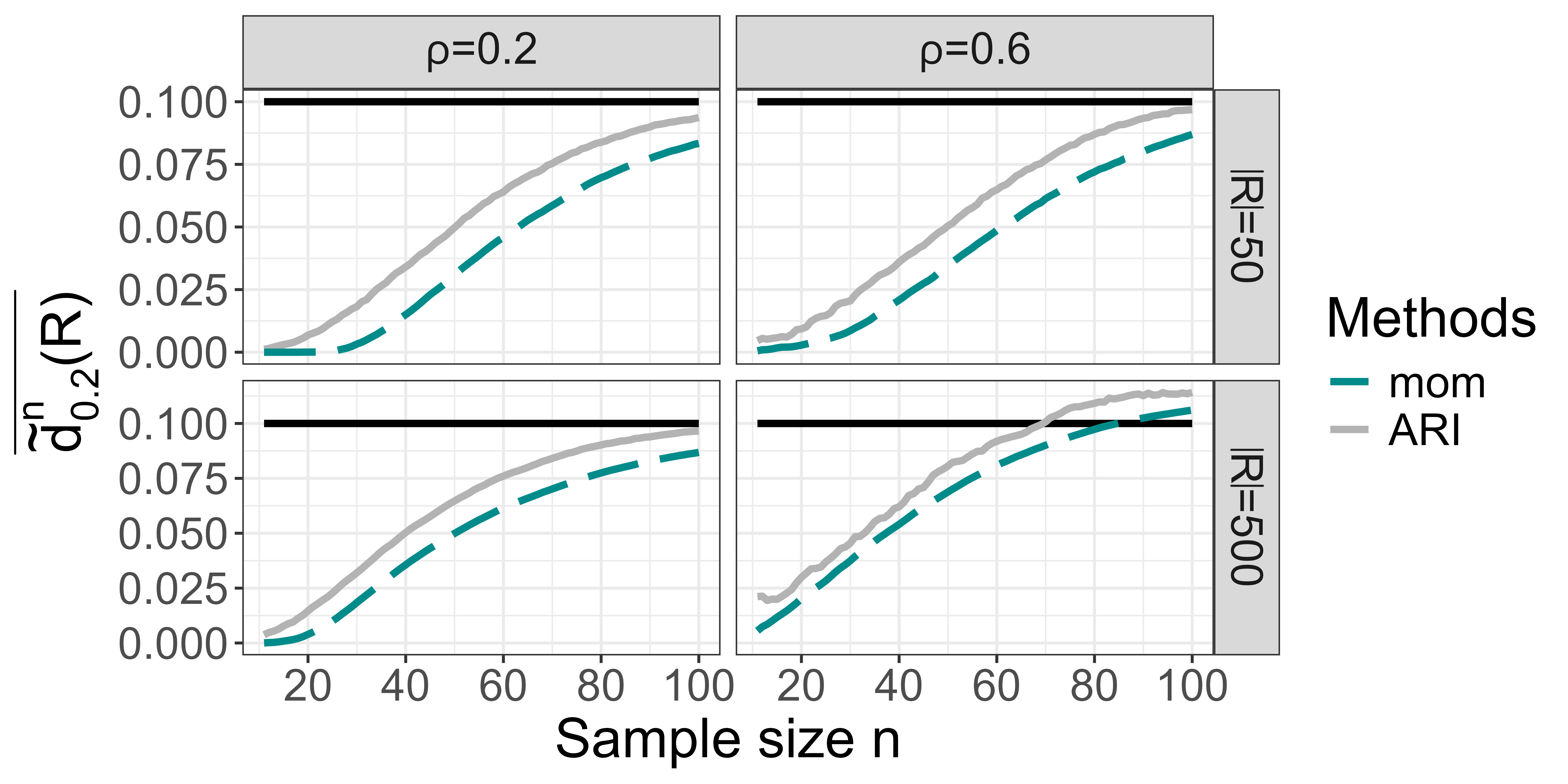}
		\\[3pt]
		\text{$\mu=0.5$, $\pi_1(R)=0.1$}
	\end{minipage}
	
\caption{Average of the $1-\alpha$ lower confidence bound $\tilde{d}_\alpha^{[n]}(R)$ for the true discovery proportion $\pi_1$ for sequentially increasing sample size $n$ and different effect sizes $\mu$, true discovery proportion $\pi_1$, level of dependency $\rho$ and size of the discovery set $|R|$.
	The anytime-valid simultaneous confidence bounds are based on the mom e-process (mom, dark dashed line).  
	The proposed procedure is compared to the all resolution inference procedure (ARI, light gray solid line) by \citet{Goeman2019}.
	The black solid line corresponds to $\pi_1$. Results are based on $1,000$ iterations and $\alpha=0.2$.}
	\label{figure:Power_rho_02}
\end{figure}

Next, we investigate the power of the proposed procedure. The procedure is considered to be powerful when the lower confidence bound $\tilde{d}_\alpha^{[n]}(R)$ is large. 
The left plot in Figure \ref{figure:Power_rho_02} displays the average of the confidence bounds across the iterations as the number of available observations increases, for $\pi_1(R)\in\{0.1,0.9\}$, $\mu\in\{0.5,1,1.5\}$, $\rho=0.2$ and $|R|=500$. 
In this simulation study, the simultaneous confidence bounds based on the ARI procedure have been at least as powerful as the proposed anytime-valid simultaneous confidence bounds. When increasing the effect size $\mu$, the average time to converge has decreased drastically. 

The right plot in Figure \ref{figure:Power_rho_02} displays the average confidence bounds as the sample size increases for $\pi_1(R)=0.1$, $\mu=0.5$ and varying $\rho$ and size of the discovery set $|R|$. 
Strong dependency between the data of the hypotheses has generally lead to slightly more powerful confidence bounds than weak dependency. 
Small discovery sets have lead to less powerful lower confidence bounds for both compared procedures than large discovery sets.

Overall, these simulations indicate that for a reasonably large effect size ($\mu\geq1$), the bounds seem to converge to the true value of the TDP in a reasonable time. Specifically, in our simulations it took on average roughly 30 observations or less. However, the anytime-valid confidence bounds took on average longer to converge to the true TDP than the bounds without anytime-validity guarantees.

\section{Case Study}
\label{sec:case_study}
The proposed method is applied to data from a task-based fMRI experiment. The analysis pipeline for fMRI data is briefly described in a general setting before the application to the observed data is discussed.

The aim of task-based fMRI studies is to find brain regions that are considered to be active when presented with a certain condition. To this end, the blood oxygenation level dependent (BOLD) signal is measured at each voxel in the brain while the participant partakes in the experiment. A voxel is a cell in the three-dimensional grid that partitions the brain during the scanning process. Typically, the BOLD signal is measured at more than $100,000$ voxels in a brain.  We only consider the case where one hypothesis is tested per voxel. Thus, in the following, $m$ refers both to the number of voxels and to the number of elementary hypotheses, leaving distinction to context. 

The first step in the analysis pipeline for fMRI data is preprocessing. Preprocessing typically includes correction of head motion, spatiotemporal filtering, slice time correction, highpass filtering and the registration of the observed brain to a standard space, among other things. 
Subsequent analysis of the fMRI data is then done in a two-step procedure. At both steps, analysis is made for each voxel $i$, $i=1,\ldots,m$. In the following, we omit the subscript $i$ for ease of notation.
In the first level analysis, the relationship between the experiment's conditions and the brain's activity is estimated for each voxel and per subject. To this end, a linear relationship between the BOLD signal and the so-called hemodynamic response is assumed. The hemodynamic response is the convolution of the condition onset time series and the hemodynamic response function (HRF). 
The BOLD signal for subject $j$ at a given voxel is modeled as \begin{equation*}
	\label{eq:first_level_analysis}
	\mathbf{y}_{j}=\mathbf{X}_{j}\boldsymbol{\beta}_{j}+\mathbf{Z}_{j}\boldsymbol{\eta}_{j}+\boldsymbol{\epsilon}_{j},
\end{equation*} where $\mathbf{y}_{j}\in\mathbb{R}^{T\times 1}$ denotes the BOLD signal measured at $T$ time points.
The design matrix $\mathbf{X}_{j}$ contains the hemodynamic response per condition and $\mathbf{Z}_{j}$ is a matrix containing confounding variables such as a baseline, trend or head motion parameters. The (linear) relationship between the hemodynamic responses (the confounding variables) and the BOLD signal is modeled by $\boldsymbol{\beta}_{j}$ ($\boldsymbol{\eta}_{j}$). Both $\boldsymbol{\beta}_{j}$ and $\boldsymbol{\eta}_{j}$ are voxel specific. The noise $\boldsymbol{\epsilon}_{j}$ is assumed to follow a multivariate normal distribution with the mean equal to zero. Pre-whitening of the data is often done since the noise is also assumed to follow an auto-regressive process \citep{Lindquist2008}. After pre-whitening, the noise $\boldsymbol{\epsilon}_{j}$ has variance $\sigma^2_{j}\mathbf{I}_T$, where $\mathbf{I}_T$ is the $(T\times T)$ identity matrix. Then, regression parameters $\boldsymbol{\beta}_{j}$ and $\boldsymbol{\eta}_j$ can be estimated using the ordinary least squares approach. If the researcher is interested in, say, the influence of the first condition on the brain's activity, the effect size $\gamma_j=\widehat{\boldsymbol{\beta}_{1,j}}/\sigma_{\widehat{\boldsymbol{\beta}_{1,j}}}$ is utilized in the subsequent analysis. The variance of the estimated regression parameter $\widehat{\boldsymbol{\beta}_{1,j}}$, denoted by $\sigma_{\widehat{\boldsymbol{\beta}_{1,j}}}^2$, is typically unknown in practice and has to be estimated as well. Since the number of observations is typically much larger than the number of predictors in the linear regression model, the estimate $\hat{\sigma}_{\widehat{\boldsymbol{\beta}_{1}},j}^2$ can be considered to be exact \citep{Thirion2017}. Therefore, we assume that the estimated effect size $\widehat{\gamma_j}=\widehat{\boldsymbol{\beta}_{1,j}}/\hat{\sigma}_{\widehat{\boldsymbol{\beta}_{1,j}}}$ has variance one.

At the second level analysis, inference is made on $\hat{\boldsymbol{\gamma}}=(\widehat{\gamma_{1}}, \ldots,\widehat{\gamma_{n}})^\top$ for each voxel, given that $n$ subjects have participated in the experiment so far. Note that $\widehat{\gamma_{1}}, \ldots,\widehat{\gamma_{n}}$ are considered to be independent and identically distributed.
The following linear model is utilized:
\begin{equation*}
	\label{eq:second_level_analysis}
	\widehat{\boldsymbol{\gamma}}=\theta\mathbf{1}  +\boldsymbol{\varepsilon},
\end{equation*} where $\mathbf{1}\in\mathbb{R}^{n\times 1}$ is a vector containing only ones. The noise $\boldsymbol{\varepsilon}$ is assumed to follow a multivariate normal distribution with a mean of zero and variance $\Sigma_1=\sigma_{B}^2+\mathbf{I}_n$, where $\mathbf{I}_n$ is the ($n\times n)$ identity matrix, corresponding to the within-subject variance and $\sigma_B^2$ denotes the between-subject variance. For each voxel $i$, the null hypothesis $H_i:\theta_i\leq 0$ is considered. 

We aim to make inference on the number of active voxels in certain brain regions, called regions of interest (ROIs). A region of interest is a cluster of neighboring voxels. One approach is to define ROIs based on the observed p-values or e-processes at each voxel. For example, we could define regions of interest as cluster of voxels with e-processes larger than a pre-defined threshold. An alternative approach is to utilize pre-specified ROIs. For example, brain regions defined in so-called atlases or brain regions defined by previous studies or meta-analyses can be used. 

Given the general analysis pipeline, we now discuss the considered fMRI experiment and the analysis of the corresponding data. Data of a semantic fMRI experiment has been provided by the Leibniz Institute for Neurobiology, Magdeburg, Combinatorial NeuroImaging Core Facility. The experiment has included a semantic and discrimination condition. For the semantic condition, word pairs have been visually presented and subjects have had to indicate whether the presented words have had the same meaning. For the discrimination condition, subjects have had to decide whether two unpronounceable letter strings were identical. The discrimination condition functions as control condition. The conditions have been presented block-wise. During one block, eight pairs of words or letter strings have been presented for 2.5 seconds each, with 0.5 seconds between each pair. In total, nine blocks of each condition have been alternately presented, with a resting period of 24 seconds between each block. More details regarding the experiment set up can be found in \citet{Bethmann2007}.

In total, $56$ participants have been observed. 
Preprocessing has been carried out using fMRIPrep 25.1.3 \citep{Esteban2019}, which is based on Nipype 1.10.0 \citep{Gorgolewski2011}. Preprocessing has included head motion correction, spatial temporal filtering, highpass filtering and slice time correction. Furthermore, the data has been registered to standard space, using the ICBM 152 Nonlinear Asymmetrical template version 2009c (MNI152NLin2009cAsym) \citep{MNIatlas1, MNIatlas2}. The preprocessed data per subject is available at {\tt OpenNeuro}, dataset ds007535 \citep{mnidata}.
First level analysis has been carried out using the {\tt{R}} package {\tt{fmri}} \citep{RfMRI}. We have been interested in the influence of the semantic condition on the activity of the brain and thus estimated the subject specific effects sizes, $\widehat{\gamma_{S-C,j}}$, $j=1,\ldots,56$, related to the contrast between the semantic condition and the control condition.
At the second level analysis, the mom e-processes corresponding to the hypotheses $H_i:\theta_i\leq 0$, $i=1,\ldots,m$ have been computed. Computation of the mom e-process requires choosing a minimum relevant effect size $\delta_{min}$, see Eq.~\eqref{eq:Perez2024_e-process}. In practice, this minimum relevant effect size is chosen based on previous studies or expert knowledge. To mimic a small prior study, we have randomly chosen three subjects and computed for each the $80\%$ quantile of the observed effect sizes $\widehat{\gamma_{S,j}}$. The smallest $80\%$ quantile has been slightly larger than $0.7$ and hence we have set $\delta_{min}=0.7$. The subjects used in the prior study have been excluded from the second level analysis. Thus, the maximum number of observations for the second level analysis has been $N=53$. We have computed the e-processes at the stopping times $n\in\{15,20,25,30,35,30,35,50,53\}$ to mimic sequentially observed subjects. The $80\%$ anytime-valid simultaneous lower confidence bounds for the number of active voxels in an ROI have been computed based on the mom process. 
We say that an ROI is of interest at time $n$ if at least 100 voxels or 10 percent of the voxels within the ROI are considered to be active at time $n$.

The ROIs have been defined in accordance with the Havard-Oxford Cortical Probabilistic Atlas (HOCPA), available at {\tt{https://www.templateflow.org}} \citep{templateflow}. We call these ROIs the HOCPA-ROIs. Each HOCPA-ROI includes voxels both in the left and right hemisphere of the brain. Therefore, we have additionally investigated the ROIs in the lateralized HOCPA atlas (HOCPAL), which splits each HOCPA-ROI into regions in the left or right hemisphere. This leads to $96$ ROIs which are nested within the original $48$ HOCPA-ROIs. Additionally, we have defined ROIs as clusters of HOCPA-ROIs based on the result of a meta-analysis of \citet{Binder2009}. They identify seven brain regions specialized for semantic tasks, namely the Posterior Inferior Lobe (PIL, including the Supramarginal Gyrus and Angular Gyrus), the Lateral Temporal Cortex (LTC, including the Middle Temporal Gyrus and posterior division of the Inferior Temporal Gyrus), the Ventral Temporal Cortex (VTC, including the Parahippocampal Gyrus and Occipital Fusiform Gyrus), the Dorsomedial Prefrontal Cortex (DMPFC, including the Superior Frontal Gyrus and Middle Frontal Gyrus), the Inferior Frontal Gyrus (IFG, including the pars triangularis and pars opercularis IFG), the Posterior Cingulate Gyrus (PCG, including the posterior division of the Cingulate Gyrus and Precuneous Cortex) and the Vendromedial Prefrontal Cortex located in the Frontal Medial Cortex (FMC).

\begin{table}[htbp]
	\caption{Regions of interest (ROI) according to the meta-analysis by \citet{Binder2009}.
		The considered ROI are the Posterior Inferior Lobe (PIL), the Lateral Temporal Cortex (LTC), the Ventral Temporal Cortex (VTC), the Dorsomedial Prefrontal Cortex (DMPFC), the Inferior Frontal Gyrus (IFG), the Posterior Cingulate Gyrus (PCG) and the Frontal Medial Cortex (FMC). The size of the ROI in number of voxels, as well as the $80\%$ anytime-valid, simultaneous lower confidence bounds for the number of active voxels after observing $n$ subjects are displayed. Bold values indicate that more than ten percent of the voxels in the ROI are considered to be active.}
		\label{tab:fMRI_resultsBinder}
		\centering
	\begin{tabular}{llrrrrrrrrrr}
		ROI &Hemi-& $|$voxels$|$& \multicolumn{9}{c}{stopping time $n$}\\
		&sphere&&15&20&25&30&35&40&45&50&53\\
		\hline
		PIL&&11,196&0&0&0&0&0&0&1&4&8\\
		LTC&&13,049&0&0&0&1&40&92&186&277&315\\
		&Left&6,262&0&0&0&0&38&87&177&266&297\\
		&Right&6,787&0&0&0&0&0&0&0&0&0\\
		VTC&&8,667&0&0&0&0&1&15&38&59&70\\
		DMPFC&&18,662&0&0&0&5&21&56&114&220&260\\
		&Left&9,708&0&0&0&4&21&56&113&220&260\\
		&Right&8,947&0&0&0&0&0&0&0&0&0\\
		IFG&&4,853&0&0&36&180&304&405&\textbf{548}&\textbf{665}&\textbf{695}\\
		&Left&2,430&0&0&36&178&\textbf{301}&\textbf{404}&\textbf{546}&\textbf{664}&\textbf{694}\\
		&Right&2,423&0&0&0&0&0&0&0&0&0\\
		PCG&&12,726&0&0&0&0&0&0&1&14&22\\
		FMC&& 1,476 & 0 & 0 & 0 & 0 & 0 & 0 & 1 & 6 & 6 \\ 
		\hline
	\end{tabular}
\end{table}
Table \ref{tab:fMRI_resultsBinder} displays the observed lower confidence bounds for the number of truly active voxels in the ROIs identified by \citet{Binder2009}. Activation has been found in all seven ROIs. Three of the ROIs have been considered to be of interest, namely the LTC, the DMPFC and the IFG. For these three ROIs, we have investigated the activation in the left and right hemisphere, respectively, to localize the activation. \citet{Binder2009} find that the activation should, in general, be stronger in the left hemisphere than the right hemisphere.
Indeed, in the three ROIs of interest, only voxels in the left hemisphere have been considered to be active.

\begin{table}[htbp]
	\caption{Regions of interest (ROI) according to the Harvard-Oxford Probabilistic Atlas which are considered to be of interest after observing 53 subjects, i.e., at least 100 voxels in the ROI are considered to be active. These ROIs are the Frontal Pole (FP), the Superior Frontal Gyrus (SFG), the Inferior Frontal Gyrus pars triangularis (IFGpt), the Inferior Frontal Gyrus pars opercularis (IFGpo), the posterior division of the Superior Temporal Gyrus (pSTG), the posterior division of the Middle Temporal Gyrus (pMTG) and the Frontal Orbital Cortex (FOC). The size of the ROI in number of voxels, as well as the $80\%$ anytime-valid simultaneous lower confidence bounds for the number of active voxels in the ROI after observing $n$ subjects are displayed. Bold values indicate that more than ten percent of the voxels in the ROI are considered to be active.}
	\label{tab:fMRI_resultsHOCPAL}
	\centering
	\begin{tabular}{llrrrrrrrrrr}
		ROI &Hemi-& $|$voxels$|$& \multicolumn{9}{c}{stopping time $n$}\\
		&sphere&&15&20&25&30&35&40&45&50&53\\
		\hline
		Brain &  & 220066  & 0  & 386  & 990  & 1922  & 2565  & 2938  & 3868  & 4675  & 4917 \\ 
		& Left & 110352  & 0  & 227  & 748  & 1628  & 2234  & 2595  & 3421  & 4160  & 4373 \\ 
		& Right & 109714  & 0  & 0  & 0  & 8  & 26  & 49  & 96  & 138  & 165  \\ 
		\hline
		FP &  & 26224  & 0  & 0  & 0  & 64  & 144  & 206  & 335  & 418  & 477  \\ 
		& Left & 12149  & 0  & 0  & 0  & 39  & 114  & 167  & 282  & 356  & 407  \\ 
		& Right & 14056  & 0  & 0  & 0  & 0  & 9  & 16  & 27  & 37  & 41  \\ 
		SFG& & 9797  & 0  & 0  & 0  & 0  & 3  & 22  & 59  & 123  & 145  \\ 
		& Left & 5289  & 0  & 0  & 0  & 0  & 3  & 22  & 59  & 123  & 145  \\ 
		& Right & 4501  & 0  & 0  & 0  & 0  & 0  & 0  & 0  & 0  & 0  \\ 
		IFGpt & & 2557  & 0  & 0  & 18  & 119  & 198  & \textbf{280}  & \textbf{366}  & \textbf{439}  & \textbf{472}  \\ 
		& Left & 1211  & 0  & 0  & 18  & 118  & \textbf{196}  & \textbf{279}  & \textbf{365}  & \textbf{438}  & \textbf{470}  \\ 
		& Right & 1346  & 0  & 0  & 0  & 0  & 0  & 0  & 0  & 0  & 0  \\ 
		IFGpo & & 2296  & 0  & 0  & 0  & 3  & 29  & 53  & 106  & 159  & 167  \\ 
		& Left & 1219  & 0  & 0  & 0  & 3  & 29  & 53  & 106  & \textbf{160}  & \textbf{167}  \\ 
		& Right & 1077  & 0  & 0  & 0  & 0  & 0  & 0  & 0  & 0  & 0  \\ 
		pSTG &  & 2269  & 0  & 0  & 0  & 0  & 9  & 26  & 74  & 138  & 140  \\ 
		& Left & 1071  & 0  & 0  & 0  & 0  & 9  & 26  & 74  & \textbf{137}  & \textbf{139}  \\ 
		& Right & 1198  & 0  & 0  & 0  & 0  & 0  & 0  & 0  & 0  & 0  \\ 
		pMTG & & 4536  & 0  & 0  & 0  & 0  & 12  & 35  & 90  & 132  & 152 \\ 
		& Left & 2220  & 0  & 0  & 0  & 0  & 12  & 35  & 87  & 128  & 147 \\ 
		& Right & 2316  & 0  & 0  & 0  & 0  & 0  & 0  & 0  & 0  & 0  \\ 
		FOC &  & 4973  & 0  & 0  & 0  & 31  & 79  & 140  & 215  & 281  & 318  \\ 
		& Left & 2657  & 0  & 0  & 0  & 29  & 77  & 137  & 207  & \textbf{270}  & \textbf{306}  \\ 
		& Right & 2316  & 0  & 0  & 0  & 0  & 0  & 0  & 0  & 2  & 2  \\ 
		\hline
	\end{tabular}
\end{table}

Table \ref{tab:fMRI_resultsHOCPAL} displays the confidence bounds for the number of active voxels in HOCPA-ROIs that are of interest after $n=53$ observations. The activity in the corresponding lateralized HOCPA-ROIs is reported as well.
At the first stopping time ($n=15$) the confidence bounds for the number of active voxels has been zero for each ROI. Having observed $n=35$ participants, one HOCPA-ROI has been considered to be of interest. After observing $n=53$ subjects, seven HOCPA-ROIs have been of interest, namely the Frontal Pole (FP), Superior Frontal Gyrus (SFG), the Inferior Frontal Gyrus pars triangularis (IFGpt), the Inferior Frontal Gyrus pars opercularis (IFGpo), the posterior division of the Superior Temporal Gyrus (pSTG), the posterior division of the Middle Temporal Gyrus (pMTG) and the Frontal Orbital Cortex (FOC).
At the last stopping time the procedure has found activation in $26$ out of $48$ HOCPA-ROIs altogether.	The largest estimated proportion of active voxels has been within the left IFGpt, where at least $38.81\%$ of voxels can be considered to be active. The confidence bounds have increased between the last two time steps in most regions, indicating that more active voxels might be found if more data is observed. 

\section{Discussion}
\label{sec:discussion}

We have proposed a novel procedure to compute carefree anytime-valid simultaneous lower confidence bounds for the TDP. 
To this end, we have utilized confidence bounds based on closed testing, where the closed testing rejection sets are determine by anytime-valid local level-$\alpha$ tests. In particular, we have considered anytime-valid local level-$\alpha$ tests based on e-processes. We have proposed a computational shortcut to make the computation of the proposed bounds feasible in settings where the number of considered hypotheses is large.

In contrast to established procedures, the proposed simultaneous lower confidence bounds allow for optional stopping. That is, the data gathering process may be stopped after any observational time point and the decision to continue may rely on the data observed so far.
For example, the decision to continue the data acquisition might be based on the sequence of computed confidence bounds. If the bounds have changed over the last few observational time points, gathering more data might lead to even more informative bounds.
Optional stopping is a desirable property in applications where data acquisition is both time and cost intensive. 
The proposed method can be applied in research areas such as genomics and neuroscience. Furthermore, it can be used for variable selection in regression models.

The proposed anytime-valid simultaneous confidence bounds are valid for any family of anytime-valid local level-$\alpha$ tests determining the set of hypotheses rejected by closed testing. In this work we focused on the family of anytime-valid local level-$\alpha$ tests $\phi_\alpha^{[n]}(E)$. 
As mentioned, $\phi_\alpha^{[I,n]}(E)$ is not carefree and can be uniformly improved by the anytime valid local level-$\alpha$ test $\tilde{\phi}_\alpha^{[I,n]}(E)\coloneq \max_{\ell\leq n}\phi_\alpha^{[I,\ell]}(E)$. The closed testing rejection set determined by $\tilde{\phi}_\alpha^{[n]}(E)$ is larger or equal to the closed testing rejection set determined by $\phi_\alpha^{[I,n]}(E)$. In turn, the proposed anytime-valid simultaneous confidence bounds are potentially tighter when using $\tilde{\phi}_\alpha^{[n]}(E)$. Thus, it would appear advantageous to utilize the family $\tilde{\phi}_\alpha^{[n]}(E)$ directly when computing the proposed bounds. However, it is not possible to determine the hypothesis intersecting exactly $h$ discoveries which is least likely to be rejected by $\tilde{\phi}_\alpha^{[n]}(E)$ based only on the e-process values of the elementary hypotheses at time $n$. This is the essential feature of $\phi_\alpha^{[n]}(E)$ which enables the computational shortcut based on Lemma \ref{lemma_shortcut}. Therefore, by utilizing $\phi_\alpha^{[n]}$ in Lemma \ref{lemma_shortcut}, we sacrifice some power but ensure that it is feasible to compute the proposed bounds even if the number of considered hypotheses is large.
An alternative computational shortcut to that proposed in Appendix A.3 can be based on \citet{Tian2023}, Algorithm 1. This algorithm needs to be adapted to the use of $\phi_\alpha^{[n]}(E)$ and the arithmetic average as e-merging function.

We have commented on the choice of the underlying e-process.  By choosing a suitable e-process, the method can be applied to any temporal dependency structure. Regardless of the utilized e-process, the proposed method allows for arbitrary dependence between the data of the considered hypotheses. The presented procedure has been derived under the assumption that the utilized e-processes are defined with respect to the same, global natural filtration. While this seems to be a sensible requirement in practical settings, it is not a necessary requirement. Indeed, the procedure can be based on e-processes with respect to any other global filtration $(\mathcal{G}_n)_{n\geq 0}$. Regardless of the considered global filtration, the proposed bounds are valid at any time point, since the anytime-valid local level-$\alpha$ tests are valid at any time point \citep[Lemma 1]{Ramdas2022}.

In a simulation study the proposed method's behavior has been investigated. In the simulations, the confidence bounds have converged to the true TDP as the number of observations increased. The average time to converge has depended on the distribution of the data, the bounds have converged faster to the true TDP for larger effect sizes. Furthermore, in our simulation study, anytime-valid confidence bounds have required a larger sample size to become tight than simultaneous confidence bounds. The observed loss of power due to anytime-validity might be less (or more) severe if the proposed procedure is based on different e-processes than those implemented in our simulation study.

We have applied the proposed method to data from an fMRI experiment in which participants partook in a semantic task. After having observed $53$ subjects, the largest lower confidence bound for the proportion of active voxels in a region has been below $40\%$. We have considered ROIs which have been identified to correspond to semantic tasks by \citet{Binder2009}. The proposed procedure has found activation in all identified ROIs. In one of these ROIs, the proportion of active voxels has been estimated to be above ten percent. The computed bounds have not converged over time, which suggests that observing more subjects might lead to larger bounds. Overall, while the case study demonstrates the usability of the proposed procedure for fMRI data analysis, it also indicates the need for specialized e-processes for the application at hand to increase the power of the proposed method.

Subsequent work therefore entails specifying guidelines for the application of the proposed procedure, focusing on the choice of the e-process in different scenarios. Furthermore, developing an e-process specifically tailored to fMRI data appears desirable. This e-process should account for the spatial, as well as temporal dependency structure in the brain and allow for possible heteroscedasticity of the data. One interesting approach appears to be the construction of e-processes by merging subject-specific e-variables. Extending the proposed procedure to different settings, such as anytime-valid simultaneous confidence bounds for knock-offs, is of interest. Lastly, closed testing based on e-values allows for simultaneous control of other error rates than the TDP \citep{Xu2025}. Thus, investigating anytime-valid simultaneous control of different error rates is another possible direction for future research.

\section*{Data availability}
The data used in the case study is available at {\tt OpenNeuro}, dataset ds007535 (HemiSpeech) \citep{mnidata}.
\section*{Acknowledgement}
	Friederike Preusse gratefully acknowledges funding by the Deutsche Forschungsgemeinschaft (DFG, German Research Foundation)- project number 281474342. The author would like to thank Dr. André Brechmann from the Leibniz Institute for Neurobiology Magdeburg, Combinatorial NeuroImaging Core Facility, for providing the data used in the case study. Sincere thanks to Prof. Thorsten Dickhaus and Dr. Monitirtha Dey at the University of Bremen for their constructive criticism regarding the first version of the manuscript. The author also thanks the editorial team and two anonymous reviewers for their constructive comments, which helped improving the manuscript.
\section*{Conflict of Interest}
The author has declared no conflict of interest.

\section*{Appendix}

\subsection*{A.1.\enspace Proof of Proposition 3.1}
We first show that $c_\alpha^{[n]}$ as defined in Proposition \ref{prop_1} is an anytime-valid upper confident bound for the number of false discoveries. Then, we argue that $\tilde{c}_\alpha^{[n]}\coloneq \min_{0\leq\ell\leq n}\{c_\alpha^{[\ell]}\}$ is an anytime-valid upper confidence bound.\\
Remember that for any $\trueP\in\mathcal{P}$, the set $T(\trueP)\equiv T$ is defined to be the index set of all true elementary hypotheses. Hence, the probability that $H_T$ is ever rejected by a given anytime-valid local level-$\alpha$ test is at most $\alpha$ by definition. Thus
	it holds for all $\trueP\in\mathcal{P}$ and for all $\alpha\in[0,1]$ that
		\begin{align*}
			\trueP(\tau(R)>c_\alpha^{[n]}(R) \text{ for any } R\in\mathcal{C}\text{ and }n\in\mathbb{N} )&\leq\trueP(T_R\in\mathcal{X}_\alpha^{[n]} \text{ for any }R\in\mathcal{C} \text{ and } n\in\mathbb{N})\\
			&\leq\trueP(T\in\mathcal{U}_\alpha^{[n]} \text{ for any }n\in\mathbb{N})\leq\alpha.
		\end{align*}
Since the event $T\notin\mathcal{U}_\alpha^{[n]}$ does not depend on the chosen discovery set $R$, the bounds are simultaneously valid for all $R\in\mathcal{C}$. 
The bounds defined by $c_\alpha^{[n]}$ may increase over time if the family of anytime-valid local level-$\alpha$ test $\phi_\alpha^{[n]}$ used to determine $\mathcal{U}_\alpha^{[n]}$ is not carefree.
Any $\phi_\alpha^{[n]}$ is either carefree or can be improved by the carefree family $\tilde{\phi}_\alpha^{[n]}$, with $\tilde{\phi}_\alpha^{[I,n]}\coloneq\max_{\ell\leq n}\{\phi_\alpha^{[I,\ell]}\}$. Utilizing a carefree anytime-valid local level-$\alpha$ test ensures that $\mathcal{U}_\alpha^{[1]}\subseteq \mathcal{U}_\alpha^{[2]}\subseteq\ldots$ and hence $\mathcal{X}_\alpha^{[1]}\subseteq \mathcal{X}_\alpha^{[2]}\subseteq\ldots$. Therefore, regardless of $\phi_\alpha^{[n]}$, it is always possible to define $\tilde{\phi}_\alpha^{[n]}$ which ensures  carefree bounds, i.e.,  $c_\alpha^{[n+1]}\leq c_\alpha^{[n]}$ for all $n\in\mathbb{N}$. Hence, for any $\phi_\alpha^{[n]}$ determining $\mathcal{U}_\alpha^{[n]}$, carefree anytime-valid simultaneous upper confidence bounds for the number of false discoveries are given by $\tilde{c}_\alpha^{[n]}$.
$\Box$

\subsection*{A.2.\enspace Proof of Lemma 3.2}
We first show that the hypothesis with index set $\K{h}$ is the least likely to be rejected by closed testing at time $n$ out of all hypotheses of size $h$ intersecting only discoveries. Then, we show that the hypothesis $\K{h}$ is not rejected by closed testing if and only if the inequality in Eq.~\eqref{eq:shortcut} holds.

The e-process value at time $n$ corresponding to the hypothesis with index set $\K{h}$ is the smallest value compared to all e-process values at time $n$ corresponding to an hypothesis of size $h$ intersecting only discoveries.
By definition of $\K{h}$, $\sum_{i\in \K{h}}\e{i}\leq \sum_{i\in I_h}\e{i}$ for all $I_h\subseteq R, |I_h|=h$. Because the arithmetic mean is used as e-merging function, \begin{equation*}
		\e{\K{h}}=h^{-1}\sum_{i\in\K{h}}\e{i}\leq h^{-1}\sum_{i\in I_h}\e{i}=\e{I_h}\quad \text{for all }I_h\subseteq R, |I_h|=h.
	\end{equation*} 
	Remember that $\phi_\alpha^{[n]}(E)$ rejects an hypothesis if its corresponding e-process value at time $n$ is above a given threshold. Hence, the hypothesis $H_{\K{h}}$ is least likely hypothesis to be rejected by $\phi_\alpha^{[n]}(E)$ at time $n$ out of all hypotheses of size $h$ intersecting only discoveries. An hypothesis $H_I$ is rejected by closed testing if all hypotheses intersecting $H_I$ are rejected by a local level-$\alpha$ test.
	Let $I^\ast\supseteq I_h$ for any $I_h\subseteq R$, $|I_h|=h$, $I_h\neq \K{h}$ and $(I^\ast\setminus I_h)\cap \K{h}=\emptyset$, such that $H_{I^\ast}$ intersects at least $h$ discoveries but does not intersect $H_{\K{h}}$. Then, at time $n$, the e-process value corresponding to $H_{I^\ast}$ is at least as large as the e-process value corresponding to the hypothesis with index set $(I^\ast\setminus I_h)\cup\K{h}$. That is, \begin{equation*}
	|I^\ast|^{-1}\left(\sum_{i\in I^\ast\setminus I_h}\e{i}+\sum_{i\in \K{h}}\e{i}\right)\leq |I^\ast|^{-1}\left(\sum_{i\in I^\ast\setminus I_h}\e{i}+\sum_{i\in I_h}\e{i}\right), \quad \text{for all }I_h\subseteq R, |I_h|=h.
	\end{equation*}
	Therefore, $\K{h}$ is the least likely hypothesis to be rejected by closed testing out of all hypotheses of size $h$ intersecting only discoveries. Thus, if $\K{h}\in\mathcal{X}_\alpha^{[n]}$ then $I_h\in\mathcal{X}_\alpha^{[n]}$ for all $I_h\subseteq R$, $|I_h|=h$.

Therefore, $c_\alpha^{[n]}(R)\coloneq\max\{|I|:I\subseteq R, I\neq \emptyset, I\notin\mathcal{X}_\alpha^{[n]}\}$ is the maximum value of $h$ such that $\K{h}\notin\mathcal{X}_\alpha^{[n]}$ if this maximum exists and zero otherwise. 

Remember that $\K{h}\notin\mathcal{X}_\alpha^{[n]}$ if there exists at least one $J\supseteq\K{h}$, $J\in\mathcal{C}$, such that $J\notin\mathcal{U}_\alpha^{[n]}$. 
Thus, for every considered value of $h$, we are interested in the intersection hypothesis $H_{J^\ast}$, $J^\ast\supseteq \K{h}$, $J^\ast\in\mathcal{C}$, which is the least likely to be rejected by the anytime-valid local level-$\alpha$ test $\phi_\alpha^{[n]}(E)$. 
To this end, it needs to hold that the e-process value corresponding to $H_{J^\ast}$ is smaller or equal to the e-process values corresponding to any other $H_J$ intersecting $\K{h}$, i.e.,
 $\e{J^\ast}\leq \e{J}$ for all $J\supseteq\K{h}$, $J\in\mathcal{C}$. Hence, if $J^\ast\in\mathcal{U}_\alpha^{[n]}$, then $J\in\mathcal{U}_\alpha^{[n]}$ for all $J\supseteq\K{h}$, $J\in\mathcal{C}$. In other words, if $H_{J^\ast}$ is rejected by $\phi_\alpha^{[n]}(E)$, then all hypotheses intersecting $\K{h}$ are rejected by $\phi_\alpha^{[n]}(E)$.\\
Note that $\K{h}\subset\K{h+1}\subset\ldots\subset \K{|R|}$ and assume that $\K{h+r}\in\mathcal{X}_\alpha^{[n]}$, $r=\{1,\ldots,|R|-h\}$, i.e., all hypotheses of size larger than $h$ intersecting only discoveries are rejected by closed testing. Therefore, $J^\ast\in\mathcal{U}_\alpha^{[n]}$ if $H_{J^\ast}$ were to intersect additional discoveries besides those in $\K{h}$. 

Define an intersection hypothesis $H_{J'}$ intersecting $H_{\K{h}}$ and $k$ non-discoveries, $J'\coloneq \K{h}\cup \BarK{k}$, where $\BarK{0}=\emptyset$ is allowed. 
Since $\phi_\alpha^{[n]}(E)$ determines the rejection set $\mathcal{U}_\alpha^{[n]}$ and the arithmetic mean is used as e-merging function, $J'\notin\mathcal{U}_\alpha^{[n]}$ iff
\begin{align}
	\label{eq:shortcut_inequality}
	\e{J'} &<\frac{1}{\alpha}\nonumber\\ 
	\Leftrightarrow	\frac{\sum_{i\in \K{h}}\e{i}+\sum_{j\in \BarK{k}}\e{j}}{h+k} &<\frac{1}{\alpha}\nonumber\\ 
	\Leftrightarrow \sum_{i\in \K{h}}\e{i}-\frac{h}{\alpha} &< \frac{k}{\alpha}-\sum_{j\in \BarK{k}}\e{j}.
\end{align} 
Remember that $H_{J^\ast}$, $J^\ast\supseteq \K{h}$ is supposed to be the intersection hypothesis which is the least likely to be rejected by $\phi_\alpha^{[n]}(E)$. Hence, we choose $k$ such that it maximizes the expression on the right side of Eq.~\eqref{eq:shortcut_inequality} and thus $J^\ast\coloneq\K{h}\cup \BarK{k^\ast}$.
Thus, if $\K{h+r}\in\mathcal{X}_\alpha^{[n]}$ for all $r=\{1,\ldots,|R|-h\}$, $J^\ast\notin\mathcal{U}_\alpha^{[n]}$ and $\K{h}\notin\mathcal{X}_\alpha^{[n]}$ if and only if the inequality in Eq.~\eqref{eq:shortcut} holds.\\

Lastly, we show that $\K{\tilde{h}(n)+r}\in\mathcal{X}_\alpha^{[n]}$ for all $r\in\{1,\ldots,|R|-\tilde{h}(n)\}$. Since $\phi_\alpha^{[n]}(E)$ can be uniformly improved by $\tilde{\phi}_\alpha^{[n]}(E)$, we can always ensure that $\mathcal{X}_\alpha^{[1]}\subseteq \mathcal{X}_\alpha^{[2]},\ldots$. Hence, it is ensured that at time point $n$, $\K{\tilde{h}(n)+r}\in\mathcal{X}_\alpha^{[n]}$ for all $r\in\{\tilde{h}(n-1)-\tilde{h}(n)+1,\ldots,|R|-\tilde{h}(n)\}$. That is, since all hypotheses intersecting more that $\tilde{h}(n-1)$ discoveries have been rejected by closed testing at time $n-1$, they are also rejected by closed testing at time $n$. Thus, we only need to consider $\K{\tilde{h}(n)+r}$ with $r\in\{1,\ldots,\tilde{h}(n-1)-\tilde{h}(n)\}$. To this end, assume that there exists an $h'=\tilde{h}(n)+r$ such that $\K{h'}\notin\mathcal{X}_\alpha^{[n]}$ and $\K{h'+r'}\in\mathcal{X}_\alpha^{[n]}$, $r'=\{1,\ldots, \tilde{h}(n-1)-h'\}$.
Therefore, the inequality in Eq.~\eqref{eq:shortcut} holds for $\K{h'}$. This contradicts the definition of $\tilde{h}(n)$ and thus $\K{\tilde{h}(n)+r}\in\mathcal{X}_\alpha^{[n]}$ for all $r\in\{1,\ldots,|R|-\tilde{h}(n)\}$.
  Hence, $\K{\tilde{h}(n)}\notin\mathcal{X}_\alpha^{[n]}$ if and only if the inequality in Eq.~\eqref{eq:shortcut} holds.
 $\Box$
 
\subsection*{A.3.\enspace Algorithm}
In this section, we propose an algorithm to compute the anytime-valid simultaneous upper confidence bound for $r$ discovery sets $R_1,\ldots, R_r$ based on Lemma \ref{lemma_shortcut} at time point $n$. Identifying the e-process values corresponding to the elementary hypotheses as discovery e-process values or non-discovery e-process values (lines 7 and 16) takes $m$ operations. Computing the cumulative sum (lines 8) requires at most $m$ operations. Determining the threshold in line 17 takes at most $2(m-1)$ operations (computing the cumulative sums and finding the maximum). Each While loop (lines 10-14 and 18-22) takes time $O(m)$. Thus, computing the upper confidence bound $\tilde{h}_{R_q}(n)$ for one discovery set $R_q$ takes linear time $O(m)$. Sorting the $m$ e-process values corresponding to the elementary hypothesis at time $n$ (line 2) takes time $O(m\log m)$. Thus, Algorithm 1 takes time $O(m\log m+mr)$, which is either $O(m\log m)$ if $r\leq \log m$ or $O(mr)$ if $r\geq \log m$.

\begin{algorithm}
	\caption{Algorithm to compute the anytime-valid upper confidence bounds $\tilde{h}_{R_1}(n),\ldots,\tilde{h}_{R_r}(n)$ for the number of true discoveries in the discovery sets $R_1,\ldots, R_r$ at time $n$ according to Lemma \ref{lemma_shortcut}.}
		\SetKwInOut{Input}{Input}
		\SetKwInOut{Output}{Output}
		\Input{
			$e_1^{[n]},\ldots, e_m^{[n]}$: e-process values at time $n$ corresponding to the elementary hypotheses\\
			$R_1,\ldots, R_r$: discovery sets\\
			$\alpha$: confidence level\\
			$\tilde{h}_{R_1}(n-1),\ldots, \tilde{h}_{R_r}(n-1)$: upper confidence bounds at time $n-1$
		}
		\hrulefill\\
			Sort the e-process values, such that $e_{[1]}^{[n]}\leq\ldots\leq e_{[m]}^{[n]}$\;
			Update the discovery sets $R_1,\ldots, R_r$ accordingly\;
		\For{$q=1,\ldots,r$}{
			Set $\tilde{h}_{R_q}(n)=0$\;
			Set $h=\tilde{h}_{R_q}(n-1)$\;
			Determine the e-process values corresponding to discoveries: $e_{[1]}^{[n]}(R_q)\leq\ldots\leq e_{[|R_q|]}^{[n]}(R_q)$\;
			Compute the cumulative sum of the e-process values corresponding to discoveries: $SR_j=\sum_{i=1}^j e_{[i]}^{[n]}(R_q)$ for $j=1,\ldots, \tilde{h}_{R_q}(n-1)$\;
			\eIf{$|R_q|=m$}{
				\While{$\tilde{h}_{R_q}(n)=0$ and $h\geq 1$}{
					\eIf{$h^{-1} SR_h<\alpha^{-1}$}{
						Set $\tilde{h}_{R_q}(n)=h$\;
					}{
						Set $h=h-1$\;
					}
				}
			}{
			Determine the e-process values corresponding to non-discoveries: $e_{[1]}^{[n]}(\Bar{R}_q)\leq \ldots\leq e_{[m-|R_q|]}^{[n]}(\Bar{R}_q)$\;
			Compute the threshold $th= \max_{0\leq k\leq (m-|R_q|)}(k/\alpha-\sum_{i=1}^k e_{[i]}^{[n]}(\Bar{R}_q))$\;
				\While{$\tilde{h}(n)=0$ and $h\geq 1$}{
				\eIf{$SR_h-h/\alpha<th$}{
					Set $\tilde{h}_{R_q}(n)=h$\;
					}{
					Set $h=h-1$\;
					}
				}
			}
		}
		
		\Return $\tilde{h}_{R_1}(n),\ldots, \tilde{h}_{R_r}(n)$
	\label{algorithm:lemma32}
\end{algorithm}

\newpage
\bibliography{references}

\end{document}